# Discontinuous Shear Thickening (DST) transition with spherical iron particles coated by adsorbed brush polymer

Georges Bossis[1,a], Yan Grasselli [1,2,b], Olga Volkova [1,c]

[1] Institute of Physics of Nice, Université University Côte d'Azur, Parc Valrose, CNRS UMR 7010, 06108, Nice, France

[2] Université University Côte d'Azur SKEMA Business School, 60 rue Dostoievski, CS30085, Sophia Antipolis, 06902, Valbonne, France

[a] Author to whom correspondence should be addressed : georges.bossis@unice.fr
[b] yan.grasselli@skema.edu
[c] olga.volkova@unice.fr

## ABSTRACT

In this work we explore the rheology of very concentrated ($0.55<\Phi<0.67$) suspensions of carbonyl iron (CI) particles coated by a small polymer. A strong DST is observed in a large range of volume fraction presenting some specificities relatively to other systems. In particular, in a given range of volume fraction, the DST transition appears suddenly without being preceded by shear thickening. The presence of a frictional network of particles is confirmed by a simultaneous measurement of the electric resistance of the suspension and of the rheological curve. Using the Wyart-Cates model we show that, increasing the volume fraction, the fraction of frictional contacts grows more and more quickly with the stress that disagrees with the prediction of computer simulations. The same kind of behavior is observed in the presence of a magnetic field with, in addition, a very strong increase of the viscosity with the magnetic field before the transition. We interpret this behavior by the interpenetration of the polymer layer under the effect of the shear stress -and of the magnetic stress- followed by the expulsion of the polymer out of the surfaces. Besides we point that, above the DST transition, we do not observe a jamming in the range of volume fraction whereas it is predicted by the W-C model. Based on the fact that in the absence of shear flow, the polymer should come back to the surface and destroy the frictional contacts we can predict an asymptotic non-zero shear rate and reproduce the experimental behavior.

## I. INTRODUCTION

The rheology of suspensions of particles is of ubiquitous importance in many industrial processes where it is needed to find a compromise between a large volume fraction of solid particles to obtain a strong material and to minimize subsequent drying keeping a low viscosity in the moulding process. Generally the viscosity of these suspensions first decreases with the shear rate (shear thinning) and then increases (shear thickening) more and more abruptly as the volume fraction of particles increases (for a review of pioneering works see Barnes [1]. If only hydrodynamic interactions between particles are present the shear rate gives a time scale but, in absence of inertia, the viscosity would not depend on it. It is only through the presence of other forces -entropic or deriving from a potential- that a dependence on the shear rate can appear. For non-Brownian suspensions adhesive Van Der Waals forces and gravity play an important role on the formation of a network of aggregated particles which can give rise to a yield stress. Increasing the shear rate will contribute to break these aggregates and to decrease he viscosity of the suspension. On the other hand, repulsive forces either electrostatic like those due to ionic layers or entropic like those coming from an adsorbed layer of polymers can prevent the aggregation. In the latter, the yield stress is decreased, and the viscosity is reduced for a given volume fraction. Generally, the decrease of the viscosity is due to deflocculating whereas the shear thickening is, on the contrary, due to the formation of transient aggregates. The qualitative explanation being that the suspending fluid imprisoned inside the aggregates behave as the solid particles and then increases the apparent volume fraction of the particles and thus the viscosity. Actually, a model based on the dynamics of aggregation/disaggregation can qualitatively represents the different rheological behaviour observed in concentrated suspensions[2]. In the extreme case of highly concentrated suspensions the shear thickening transition can manifest by a sudden jump of stress at a given shear rate in an imposed ramp of shear rate or in a sudden decrease of shear rate in a controlled stress experiment. This sudden phenomenon is called discontinuous shear thickening (DST). To our knowledge H. Freundlich[3] was the first to present an experiment clearly showing a DST transition on a paste made of quartz particles in water. More recently Hoffman [4] conducted a systematic study of the rheology of suspensions of monodisperse PVC spheres with a diameter in the range 0.4-1.3µm and volume fraction larger than 50%. The use of diffraction of white light during the experiment clearly demonstrated that the transition was associated with the rupture of a layered structure made of particles hexagonally packed and sliding over each other. This behaviour was recovered on monodisperse suspensions of smaller particles: d=200nm [5]. On the other hand this abrupt shear thickening was also observed in moderately polydisperse suspensions of latex particles by Laun *et al.*[6] and it was demonstrated, by neutron scattering in [7,8] that it happens in the absence of a layered pattern preceding the transition. Using dichroism measurements, d'Haene *et al.* [9] have observed on suspensions of PMMA sterically stabilized, that above the critical stress, the relaxation of the structure was much longer and deduced the presence of large clusters spanning the cell. Furthermore polydisperse suspensions particles of irregular shape like corn-starch [10] or acicular calcium carbonate [11], or gypsum [12] which cannot be supposed to flow in regular planes also show this jump of viscosity. The DST transition can happen as well for particles sterically stabilized in non-polar solvents like PMMA in aliphatic hydrocarbon [13] also in di-octyl phthalate [14] or stabilized by electrostatic layers in polar solvent [3], for quartz in water[15,16] for silica in water; [8,17] for silica in tetrahydrofuran; [18] for polystyrene in water. The onset of the transition is ruled by the competition between the shear forces and the repulsive forces which prevent the surfaces to come in contact and to experiment friction forces. By varying the pH in suspensions of silica or alumina at a constant salt concentration, Franks *et al.*[19] have shown that the increase of the magnitude of the repulsive force was increasing the critical shear stress of the DST transition. The sudden contact between particles is believed to provoke the formation of a network of particles acting like a solid skeleton able to support the stress through elastoplastic contacts. This network is a transient one and its rupture and reformation with the strain manifests through huge fluctuations of the stress if the shear rate is imposed [16] or of the shear rate if the stress is imposed [20] as also pointed by other authors [8,9,18]. The presence of frictional forces was confirmed experimentally through the presence of a positive normal stress proportional to the shear stress [10,20–23] which expresses the force exerted on the upper plate of the rheometer by quasi solids aggregates trying to rotate in a confined space. Also shear reversal experiments demonstrated that even in the continuous shear thickening domain, the elastic forces were dominant upon hydrodynamic ones [24]. In the usual rheometric cells, there is a free surface and the particle pressure generated by the shear stress above the transition will push them outside the fluid phase; nevertheless there are maintained inside the fluid as long as the capillary pressure is larger than the particle pressure [23,25].

Besides the experiments, numerical simulation of the trajectories of an assembly of particles is a precious tool to correlate the macroscopic observations of the rheological behavior to the spatial reorganisation of the particles under shear. Stokesian dynamics allows to properly take into account hydrodynamic forces in concentrated suspensions and in particular the lubrication forces which play a crucial role to prevent the contacts between the surfaces of the particles. A simulation including lubrication forces and frictional ones was realized by Seto *et al.* [26] and Mari *et al.* [27] on concentrated suspensions. In their model the cut-off of the lubrication forces was taken at $10^{-3}$ a (a is the radius of the particles) and at this distance, the contact forces were introduced through normal, $k_n$, and tangential $k_t$ spring stiffness with the Coulomb criteria for the tangential force : $F_t \leq \mu F_n$ where $\mu$ is the friction coefficient. It is also worthwhile to note that the presence of periodic boundary conditions prevents the dilatancy of the suspension. The simulation, made at imposed shear rate, actually shows a DST transition above a given volume fraction. The stress corresponding to the beginning of the shear thickening remains quite independent of the volume fraction-except close to the jamming one [28] - as expected if it results from a balance of repulsive to shear forces. This independence was also found experimentally in [8,14,20,22,23,29–31]. The most important observation was the correlation between the change of the fraction of frictional contacts $f(\sigma)$ with the giant fluctuation of stress close to the DST transition showing that the transition was mostly not structural but related to the contact between the particles. The function $f(\sigma)$ had a sigmoid shape and was independent of the volume fraction. In imposed stress simulations Singh *et al.* [28] have compared their results to the predictions of a model of M. Wyart and Cates [32] based on a jamming volume fraction $\Phi_j^\mu < \Phi_j(\sigma) < \Phi_j^0$ which can change from a lower bound depending on $\mu$: $\Phi_j^\mu$ to the maximum packing fraction of frictionless spheres $\Phi_0$ according to a linear equation: $\Phi_j(\sigma)=f(\sigma) \Phi_j^\mu+(1- f(\sigma))\Phi_j^0$. The model was able to reproduce the numerical data both for the viscosity, the normal stress difference and the particle pressure. Nevertheless, using this model with the values obtained for the dependence $\Phi_j^\mu (\mu)$ and different prefactors, Lee *et al.* [33] did not succeed to represent properly their experimental data on silica spheres with different coating, even with an overestimated value ($\mu=1$) of the friction coefficient. In this paper we are also comparing the prediction of this model to experimental data obtained on a suspension of iron microparticles coated by a polymer brush. Polymer brushes make very efficient coating to prevent dry friction between the particles [34]. Polymer brushes as surfactant, together with the use of mineral undeformable particles, provide a system more reliable than soft particles as shown in a recent work of Le *et al.* [35] proofing that the interaction of the solvent with the surface of the particles modifies the interparticle force and rules the DST behavior. In addition, the use of iron spheres provides a double interest: first we can measure its conductivity and we expect to observe a change when frictional contacts occur during the DST transition, second, we can add a supplementary stress through the application of a magnetic field. In this way we have already shown that it was possible to trigger the DST transition [36]. Other possibilities to trigger the transition are to add vibrations which contribute to break the force chains [37] or to use a coating of microparticles with temperature responsive polymers to modify the interparticle forces [38].

In this work we shall rather use this possibility to deepen our understanding of the DST transition.

In the first section we shall present the suspension used and the determination of its theoretical maximum flowing volume fraction $\Phi_0$. In the second part we shall look at the dependence of stress-shear rate curves versus the volume fraction and will see to which extent it is possible to reproduce them with the model of Wyart and Cates. In the third part we shall discuss these results with the help of the measurement of the electric resistance of the sample and propose some modifications of the model to explain the absence of jamming at high volume fraction. At last, looking at the change of DST with the magnetic field we shall see if it is possible to get a coherent view of these experimental data.

## II. MATERIALS

The particles we used are made of carbonyl iron obtained from BASF (grade HQ); they have a density $\rho$=7.7g/cm3 measured with a gas pycnometer and are currently used for making magnetorheological suspensions. Their size distribution was obtained with the help of several MEB images for a total of 2300 particles analysed with ImageJ. From this size distribution, the first moment is the mean radius: $M_1$= <a>=0.296µm and the standard deviation is $\sigma_{std}$= 0.15µm. A representative MEB image is shown in Fig 1a and the experimental size distribution in Fig 1b together with its fit by a lognormal distribution for the density of probability to find a radius $a_i$ in a class i of thickness 0.05mm:

$$P(x) = \frac{1}{x\sigma\sqrt{2\pi}} \exp\left[-\frac{(\ln x + 0.5\sigma^2)^2}{2\sigma^2}\right] \quad (1)$$

Here x =a/<a>. The parameter of the fit is $\sigma$=0.547. For a suspension of monosized frictionless hard spheres the maximum flowing volume fraction is the well-known random close packing $\Phi_{RCP}$=0.637. For a polydisperse suspension we use an expression based on the three first moments of the distribution[39]:

$$\frac{\Phi_0}{1-\Phi_0} = \frac{M_3}{M_1 M_2^2} \frac{\Phi_{RCP}}{1-\Phi_{RCP}} \quad \text{where } M_k = \int_0^\infty a^k P(a)da \quad (2)$$

Taking the moments $M_k$ from the experimental size distribution gives $\frac{M_3}{M_1 M_2^2} = 1.19$ and from Eq.(2) we find $\Phi_0$=0.676 for the maximum flowing fraction. Another way to obtain $\Phi_0$ is to make a correspondence between the lognormal distribution and a bidisperse suspension characterized by the two sizes of the particles and their relative proportion. This relation implies to preserve the same mean radius, and the same polydispersity and skewness for the lognormal and the bidisperse distribution [40]. In our case we find respectively for the large and small particles $a_s$=0.223µm, $a_L$=0.715mm and a proportion of large particles: $X_L$=0.149. An analytical expression for the random close packing of a bidisperse suspension was given in [41]. (cf their Eq.(16) with $\beta^{rcp}$=0.2 and $\beta^{rlp}$=0.16). Using their expression for small values of u=$a_s$/$a_L$, which is our case, we obtain $\Phi_0$=0.683. This value is somewhat larger than the one obtained from the moments of the size distribution. We shall see in the next section that it corresponds better to the value obtained from a fit of the low shear rate viscosity. Using the MEB pictures, we do not consider the thickness of the polymer layer which prevents the aggregation of the particles. This polymer is a superplastifier molecule whose commercial name is Optima 100 made of a short polyethylene oxide (PEO) chain (in average 44 O-$CH_2CH_2$ groups) and a diphosphonate head with sodium counter ions. As in a preceding work where we have used it with calcium carbonate particles we shall name it PPP44 [42]. It is the phosphonate head negatively charged which binds electrostatically with the iron surface. In all the suspensions the mass of PPP44 used was 2mg/g of iron which is slightly larger than the concentration corresponding to the inflexion of the adsorption isotherm marking the realization of the first layer of polymer on the surface of the particles. The thickness of the layer can be approximated by the gyration radius of the polymer in a good solvent which is $d=b.P^{3/5}$ with b=0.526nm the Kuhn length of the PEO group and P=44 the number of monomers; we obtain d=5.1nm. The third moment of the distribution is proportional to the volume of the solid, so taking $(a+\delta)^3$ instead of $a^3$ we obtain for the real volume fraction of the solid phase:

$$\Phi_{eff} = \frac{1}{1+\frac{M_3\ (1-\Phi)}{M_{3\delta}\ \Phi}} \qquad (3)$$

where $M_3$=0.0642 and $M_{3\delta}$ =0.066 are respectively the moments of the experimental distribution based on $a^3$ and $(a+\delta)^3$. The different volume fractions are calculated from the density of the iron particles and of the suspending fluid and corrected with the help of Eq.(3) when they are used in the rheological models.

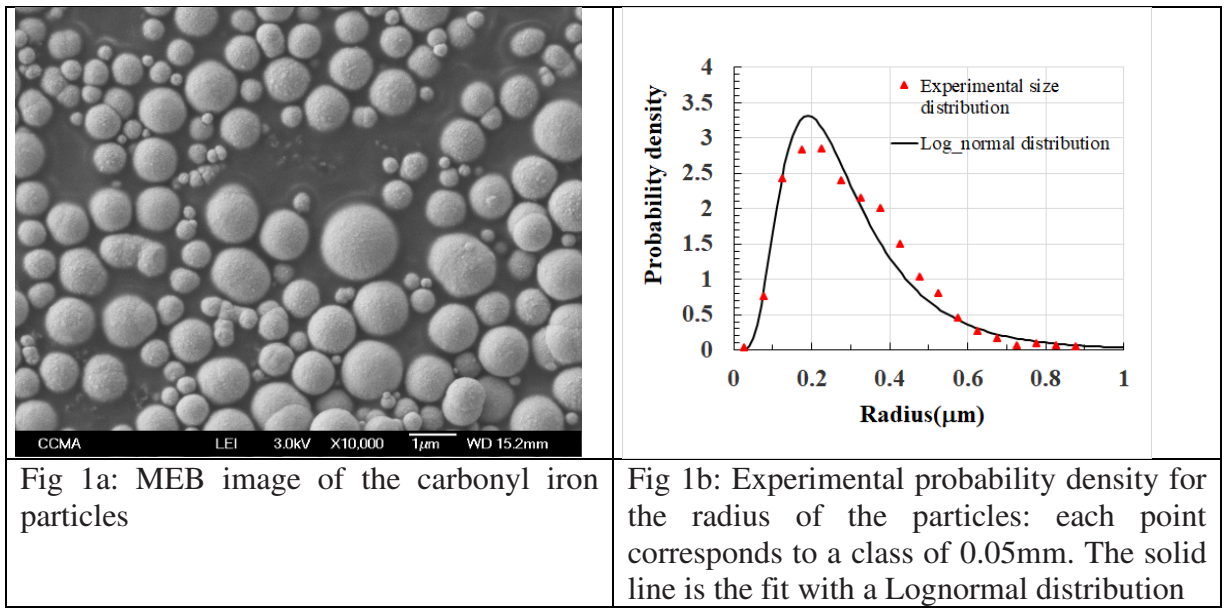

| Fig 1a: MEB image of the carbonyl iron particles | Fig 1b: Experimental probability density for the radius of the particles: each point corresponds to a class of 0.05mm. The solid line is the fit with a Lognormal distribution |
|---|---|

Finally the suspending liquid is a mixture of ethylene glycol and water (respectively 85% and 15% in mass) whose composition was chosen to minimize the evaporation rate. The viscosity of the suspending fluid at 20°C was $\eta_f$=11.8mPa.s.

## III. RHEOLOGY IN THE ABSRNCE OF MAGNETIC FIELD
### A. Experimental results

The rheogram of the suspension was obtained with an imposed stress rheometer MCR 502 from Anton Paar. Most of the experiments were realized with a plate-plate geometry where both plates were covered with sandpaper of granulometry 40 μm to avoid slipping on the walls. The usual gap was 1mm and we have to point that below 0.5mm we have noticed that the critical shear stress was decreasing. In all the experiments we take care that there is no spilling of the suspension, which can be the case at high stresses and/or high shear rates; if it was the case we have used the cylindrical Couette geometry with a small gap in order to prevent the migration of the particles from the higher shear rate domain close to the internal wall towards the external one.

In plate-plate geometry the shear rate is not constant and the stress versus shear rate curve must be corrected using the Mooney-Rabinovitch equation:

$$\tau = \frac{\tau_a}{4}\left[3 + \frac{\dot{\gamma}}{\tau_a}\frac{d\tau_a}{d\dot{\gamma}}\right] \quad (4)$$

where $\tau_a$ is the shear stress given by the software of the rheometer. We have plotted in Fig.2 The stress versus shear rate for an experiment made at a volume fraction Φ=0.64 in plate-plate geometry (black curve) and in cylindrical geometry (red curve). The first remarkable thing is that, in both geometries, we observe a sudden decrease of the shear rate by an order of magnitude at a critical point ($\dot{\gamma}_c, \sigma_c$) and that above the critical point the shear rate oscillates about a constant value in the plate-plate geometry or with a slight increase in the case of the cylindrical geometry. It is clear that for this volume fraction we do not observe a second branch with a stable flow up to the maximum stress we were able to use we; here 3800 Pa in cylindrical geometry and only 1300 Pa in plate-plate geometry because, as can be seen in the Fig.2 it ends up with an expulsion of the liquid.

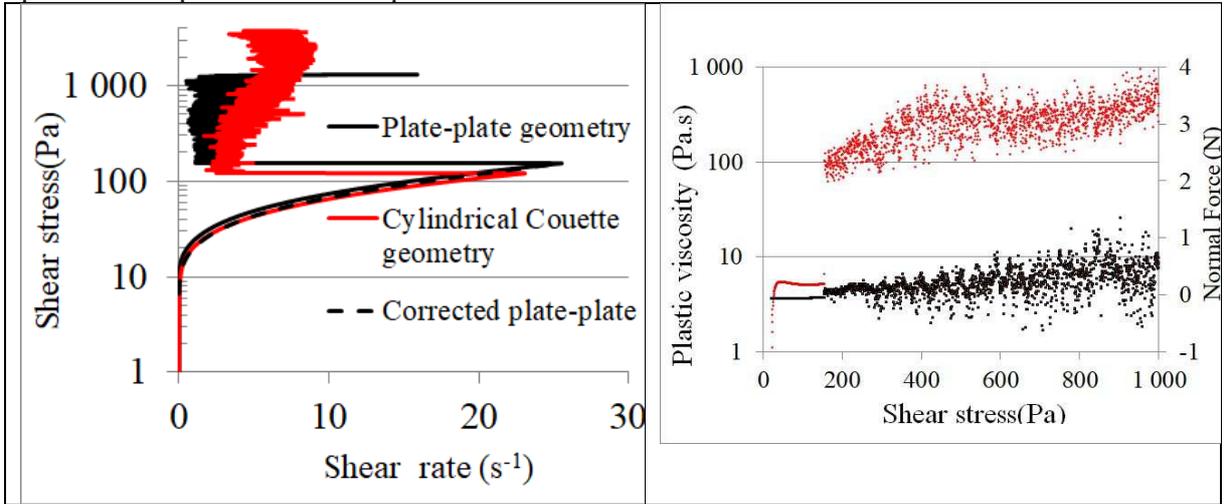

| Fig.2 Volume fraction Φ=0.64. Solid black line:Plate-plate geometry; Dotted black line: correction with Eq.(4). Solid red line: Cylindrical Couette geometry | Fig.3 Plastic viscosity (red dots, left scale) and Normal Force (black dots, right scale) in plate-plate geometry; Volume fraction Φ=0.64 |
|---|---|

The dotted line represents the application of Eq.(4) to the experimental curve below the critical stress. The experimental curve is first smoothed to calculate the derivative. We see that the corrected curve is lower that the experimental one and well follow the one obtained in cylindrical geometry below 10s$^{-1}$. After it begins to get closer to the initial curve, because the derivative $\frac{d\tau_a}{d\dot{\gamma}}$. is not constant but slightly increases as the suspension shear thickens when it approaches the jamming point. If the experimental curve was following a pure Bingham law with a yield stress $\tau_y^a$ and a plastic viscosity $\eta_p$, then the corrected curve would have a true yield stress $\tau_y = \frac{3}{4}\tau_y^a$ and the same plastic viscosity. Then in this case the true critical stress obtained in plate-plate geometry would be simply $\tau_c = \tau_c^a - \tau_y^a/4$. Nevertheless even a small shear thickening can give a quite different result, so for all the measurements made in plate-plate geometry we have used Eq.(4) to get the critical stress. In fig.3 we have plotted the normal force and the plastic viscosity: $\eta_p = (\tau-\tau_y)/\dot{\gamma}$ versus the stress. Below the jamming stress, the viscosity is almost constant: $\eta_p$=5.2±0.2 Pa.s and it jumps to $\eta_p$=90±20 Pa.s at the jamming point; at the same time, the normal force passes from slightly negative($F_n$= -0.06±0.01N) to positive ($F_n$=-0.08±0.02N). Above the critical point both values increase with the stress and fluctuate a lot; nevertheless, as previously noted [21,23], the average stress on the upper plate : $F_n/\pi R^2$ remains proportional to the shear stress with, in our case, a coefficient of 0.29±0.03.

Another important observation is that using different gaps in plate-plate geometry between 0.5mm and 1.5 mm give the same result, which indicates that there is no noticeable slip on the plates. In the following we have used plate-plate geometry with a gap of 1mm or, when high shear rates are used for Φ<0.62, a cylindrical geometry. In Figs 4a and 4b we have gathered the shear stress versus shear rate curves obtained for volume fractions between Φ=0.53 and Φ=0.67. The curve for Φ=0.62 is reported on both graphs : it is the volume fraction above which the critical stress of jamming $\sigma_c$ steadily increases until the discontinuous jamming transition disappears at Φ=0.53; also for Φ≤0.62 the DST transition is accompanied by a sudden decrease of the shear rate followed by strong oscillations which are the signature of an instability described in detail for corn-starch suspensions [43] and explained by introducing the inertia of the rotating tool [42,44].

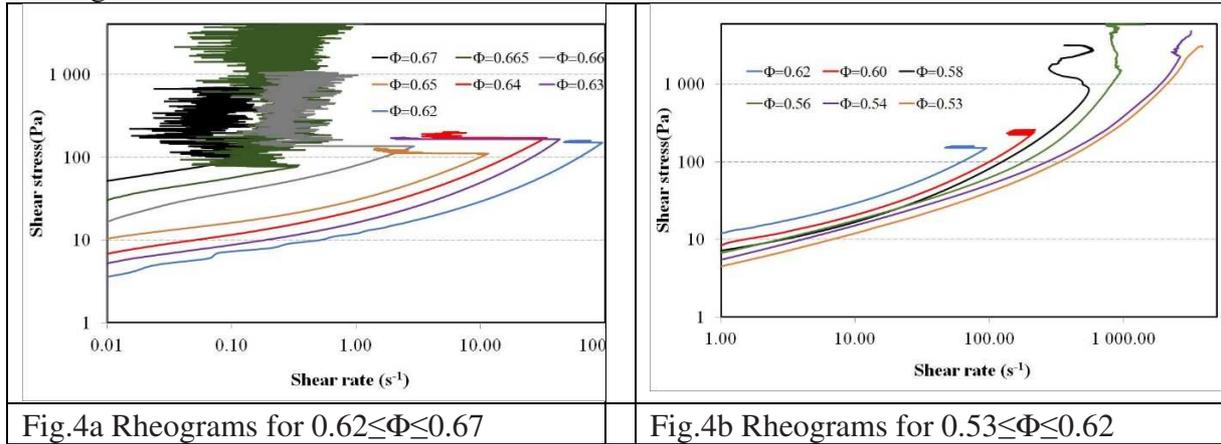

| Fig.4a Rheograms for 0.62≤Φ≤0.67 | Fig.4b Rheograms for 0.53≤Φ≤0.62 |

Another point that we want to emphasize is the fact that, even at the highest possible fraction: Φ=0.67, after the DST transition we never observe a return to zero of the shear rate as predicted by the model of Wyart et Cates [32] but rather an oscillating regime whose average value remains practically constant during the increase of the shear stress. This regime is shown in Fig.4a for Φ=0.66,0.665,0.67 and in Fig.2 for Φ=0.64. A similar behaviour was reported for other kinds of suspensions e.g. corn-starch [43,45], submicronic PMMA suspensions stabilized with grafted polymers [9,14], polystyrene particles of diameter 0.3μm [21], silica [15] and alumina particles of micronic size in water at different pH. This behaviour with a shear rate fluctuating around a constant value above the critical shear stress in stress controlled experiments seems quite generic and can't be explained by the Wyart-Cates model. We shall come back to this point at the end of this section. In the following figures we have plotted the plastic viscosity obtained from a fit of the linear part of the curve by a Bingham law $\eta(\dot\gamma) = \tau_y + \eta_{pl}\dot\gamma$ versus the effective volume fraction given by Eq.(3).

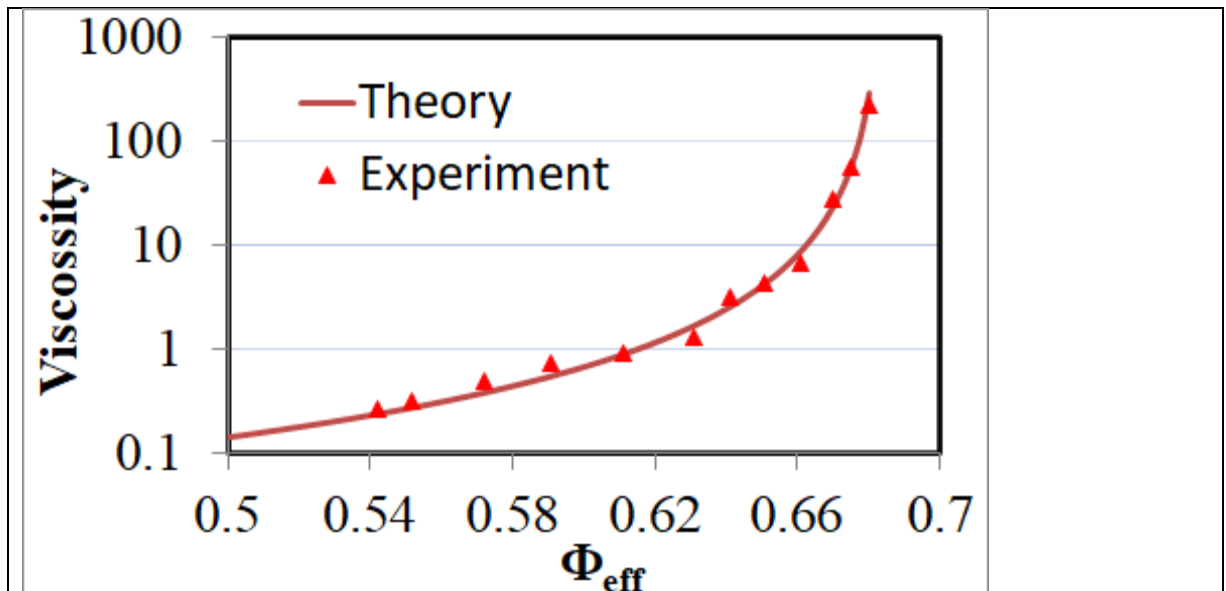

Fig.5 Plastic viscosity versus the volume fraction: $\eta(\Phi_{eff}) = A(1 - \frac{\Phi_{eff}}{\Phi_{0eff}})^{-2}$ With A=0.01 and $\Phi_{0eff}$=0.684

The value $\Phi_{0eff}$=0.684 given by the fit is very close to the prediction obtained from the use of the equivalence with a bidisperse suspension (cf section Materials) and we shall keep it for the analysis of the rheological model.

Before trying to explain the experimental curves with the help of a rheological model we also want to mention that all the curves presented are taken during the first rise of stress or shear rate (in rate control experiments) after loading the suspension. As can be seen in Fig 6, the

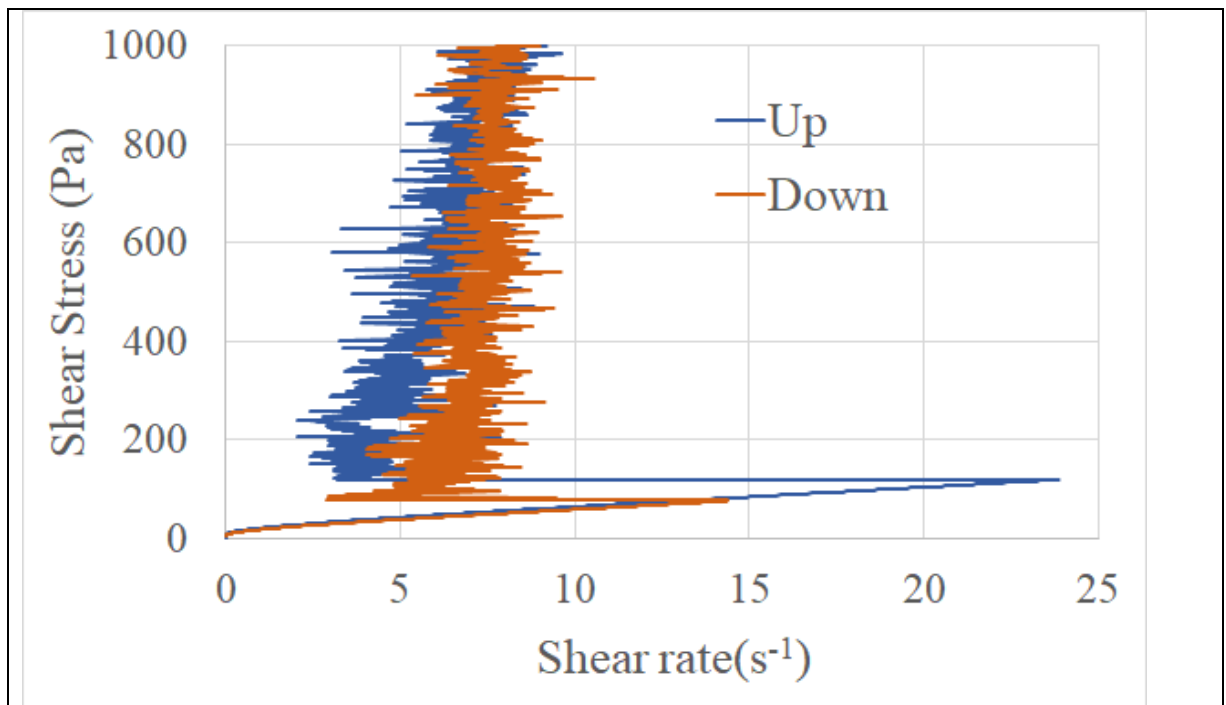

Fig.6 Hysteresis of the stress versus shear rate; the blue solid line is the ascending ramp of stress and the brown dashed line the descending ramp

descending curve (brown dashed line) shows a hysteresis and falls on the ascending one at a lower stress, but after that remains on the ascending one, showing that the suspension has recovered its equilibrium state. This hysteresis can be much more important when the DST transition occurs at high stresses as is the case at volume fraction lower than 0.62. In this case a second ramp of stress with the same suspension gives a higher viscosity on the descending branch, which is the signature of an irreversible aggregation induced by the preceding high stress. The initial state can nevertheless be recovered by a pre-shear at an intermediate stress

The fact that the DST transition is provoked by a percolation of frictional contacts between particles has been demonstrated by numerical simulation [27] but to our knowledge there is no experimental demonstration of the correlation between the percolation of contact forces and the DST transition. In our suspension we are using coated iron particles and this coating increases the resistivity of the suspension. If, on the other hand, the DST transition is related to the formation of a percolated network of frictional contacts, it means that the coating has been removed and the transition should manifest through a decrease of the conductivity of the suspension. The resistance between the two plates or between the outside cylinder and the bob was measured using a comb of thin conductive wires rubbing on the shaft. We have presented in Fig.(7) and Fig.(8) the change of the resistance of the suspension associated with the DST transition respectively for a volume fraction $\Phi=0.64$ in the domain where there is a strong decrease of the shear rate and for $\Phi=0.55$ which is close to the lower limit of the DST transition. Both measurements are made in an imposed shear rate ramp in cylindrical Couette geometry where high stresses are accessible without the expulsion of the suspension contrary to plate-plate geometry. At $\Phi=0.64$ we have imposed a ramp of shear rate, and before the definitive transition at $\dot{\gamma}=24$ s$^{-1}$, we see a transient exploration of the high stress domain accompanied by a sudden drop of the resistance which is the negative footprint of the stress jump. The drop of the resistance and the jump of shear stress are very well correlated in both cases. The effective surface of contact between the particles is difficult to quantify because a part of the conductivity can be due to tunnel effect [46]. Nevertheless, it is clear that this sudden decrease of resistance is an experimental proof of the formation of a percolation network of frictional contacts between the particles.

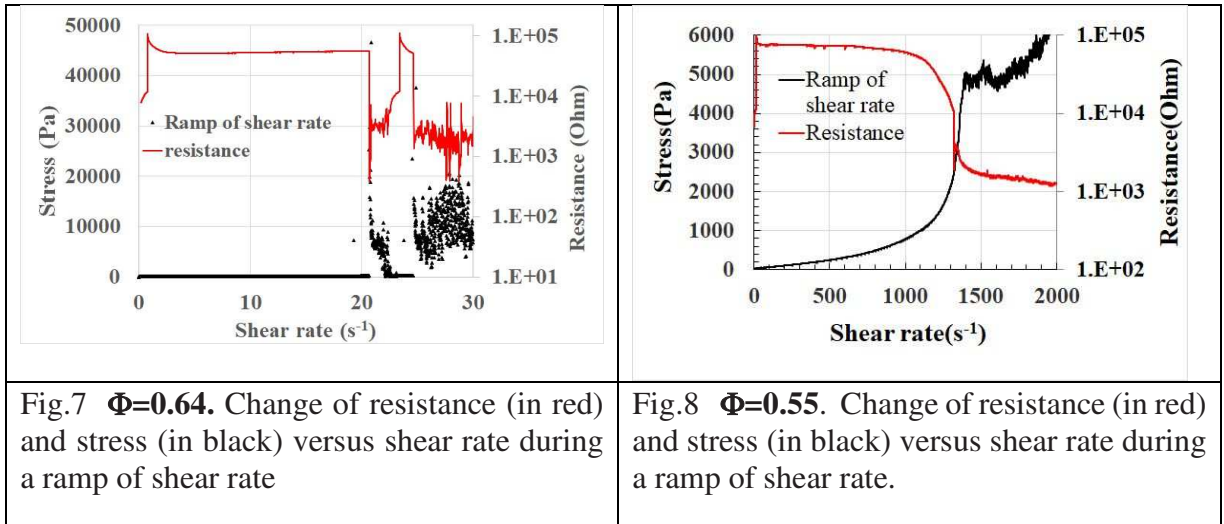

| Fig.7 $\Phi=0.64$. Change of resistance (in red) and stress (in black) versus shear rate during a ramp of shear rate | Fig.8 $\Phi=0.55$. Change of resistance (in red) and stress (in black) versus shear rate during a ramp of shear rate. |
|---|---|

For a lower volume fraction, $\Phi=0.55$, the change of resistance as well as the change of stress during a ramp of shear rate is much softer but in total is of the same order of magnitude. We can also note that at the beginning of the shear there is an increase of the resistance, which is due to the resuspension of the particles and to the destruction of a fragile network which was formed at the bottom of the cylinder in the presence of sedimentation. Besides the experimental proof of the presence of frictional contacts between particles this experiment also shows that the formation of this network can be either progressive at the lower volume fraction or very abrupt at higher volume fraction. We shall come back to this point in the next section

## B. Comparison with Wyart-Cates model

The discontinuous shear thickening transition is characterized by a point in the rheogram $\sigma = f(\dot{\gamma})$ where the derivative $\frac{d\dot{\gamma}}{d\sigma} = 0$. As $\dot{\gamma} = \sigma/\eta(\sigma)$, taking the derivative gives the condition: $\beta = \frac{dLog(\eta)}{dLog(\sigma)} = 1$ which is often used in the plot $\eta = f(\sigma)$ in Log-Log scale to characterize the DST transition. The fact that the viscosity grows with the applied stress is known as shear thickening if $\beta < 1$ and, as discussed in the introduction is related, in the case of suspensions of solid spheres, to their-transient or irreversible -aggregation. This aggregation can happen when, during the trajectory of a pair of spheres almost at contact, the shear force, which scales as $\sigma/\pi a^2$, becomes larger than the repulsive force which prevents the particles to aggregate. In this case the surfaces of the particles can come into contact and experiment a solid friction which will contribute to form quasi solid clusters of particles, thus increasing the apparent volume fraction and so the viscosity. If these clusters connect each other through dry friction and percolate throughout the cell, we end up with a jamming transition at a volume fraction $\Phi_j$ which will depend on the friction coefficient [47]. Among the repulsive force between two approaching spheres is the lubrication one which diverges as $\delta v_r / \varepsilon$ where $\varepsilon$ is the separation distance between the surfaces of two spheres and $\delta v_r$ their relative radial velocity. For smooth hard spheres the lubrication force will always prevent the surfaces of the particles to come into dry contact and the viscosity will only diverge when $\varepsilon \to 0$ for all pairs of spheres included in a percolating network at random close packing: $\Phi_{RCP} = 0.64$ if the structure remains isotropic. If the surface of the particles present asperities, the lubrication force between the surfaces of the asperities will prevent the relative translation or rotation of the particles and conduct to a jamming transition at lower volume fraction exactly as in the case of solid friction between particles as demonstrated by numerical simulation [48].

In the model of Wyart and Cates [32] it is the breakup of the lubricating film when the force generated by the particle pressure, P, overcomes the magnitude of the repulsive force, F*, ($\frac{P}{P^*} \propto \frac{\sigma}{\sigma^*} > 1$ with $P^*, \sigma^* \sim F^*/a^2$) which produces frictional contacts and a divergence of the viscosity at a volume fraction lower than $\Phi_0$. They introduce the volume fraction $\Phi_j(\sigma)$ above which the viscosity diverges. It is then only at infinite stress ($\sigma_r = \frac{\sigma}{\sigma^*} \to \infty$) that the viscosity can diverge at the lower jamming volume fraction called $\Phi_j^\mu$. In granular materials the minimum volume fraction which can support a stress is known as the random loose packing, $\Phi_{RLP} \sim 0.55$-$0.56$ for monodisperse spheres, so it seems reasonable to think that $\Phi_j^\mu \# \Phi_{RLP}$. When the pressure decreases, the proportion, f, of frictional contacts should decrease and, as in granular materials, the jamming volume fraction $\Phi_j(\sigma)$ should increase[47]. The following linear relation is taken in the W-C. model:

$$\Phi_j(f) = \Phi_j^\mu f + \Phi_0(1-f) \text{ where } f = f(\sigma_r) \text{ with } 0 < f < 1 \quad (5)$$

The divergence of the relative viscosity at $\Phi_j$ is then supposed to follow the same law than for frictionless spheres with $\Phi_j(f)$ replacing $\Phi_0$:

$$\eta_r(\Phi) \propto \left(\Phi_j(f) - \Phi_{eff}\right)^{-2} \quad or \quad \eta_r(\Phi) \propto \left(1 - \frac{\Phi_{eff}}{\Phi_j(f)}\right)^{-2} \quad (6)$$

The second expression is equivalent, close to the divergence, but is more usual in rheology so we shall keep it in the following. The power -2 was shown to well represent experimentally the divergence of the viscosity at the vicinity of the jamming volume fraction [49,50] and is also compatible with numerical simulations whatever the value of the friction coefficient [28,51,52]. When the stress and so $f(\sigma_r)$ increases, $\Phi_j(f)$ decreases and can reach the actual value of $\Phi$ if $\Phi_j^\mu < \Phi < \Phi_0$, then the viscosity diverges and the flow should stop. In a range of volume fraction below $\Phi_j^\mu$ there is still a domain of stress where the DST transition subsists but do not lead to a jammed situation. A phase diagram in the plane $(\sigma,\Phi)$ illustrating these different behavior can be found in A.Singh et al [28]. In the W-C. paper the function $f(\sigma_r)$ was chosen arbitrary as $f(\sigma_r) = 1-\exp(-\sigma_r)$. In a recent paper R. Radhakrishnan et al. [52] the authors have calculated the function $f(\sigma_r)$ from the distribution of normal forces between particles: $P(\theta)$ with $\theta = F_n/\langle F_n \rangle$ and

$$f(\theta^*) = \frac{\int_{\theta^*}^{\infty} P(\theta)d\theta}{\int_0^{\infty} P(\theta)d\theta} \qquad (7)$$

The average value of the normal force should be proportional to the applied stress $\sigma$ so $\theta^* = \frac{\lambda\sigma^*}{\sigma} = \lambda/\sigma_r$ where $\lambda=1.85$ was a proportionality constant found numerically to be independent of the volume fraction. We have plotted in Fig.(9) a curve representing the values they have obtained by numerical simulation in the low friction limit ($\mu=10^{-4}$) and a fit by the function:

$$f(\sigma_r)=e^{-\left(\frac{\lambda}{\sigma_r}\right)^q} \text{ with } \lambda=1.712 \text{ and } q= 1.163 \qquad (8)$$

The parameter $\lambda$ shifts the curves since it scales $\sigma_r=\sigma/\sigma_c$ and should not depend too much on the volume fraction since the characteristic magnitude of the repulsive force does not depend on the volume fraction; the parameter $q$ modifies the sharpness of the transition. On the other hand, as shown in this last paper, the function $f(\sigma_r)$ is quite insensitive to the value of $\mu$. Eq.(8) for $f(\sigma_r)$ was previously used [28,53] to represent the data obtained from numerical simulations.

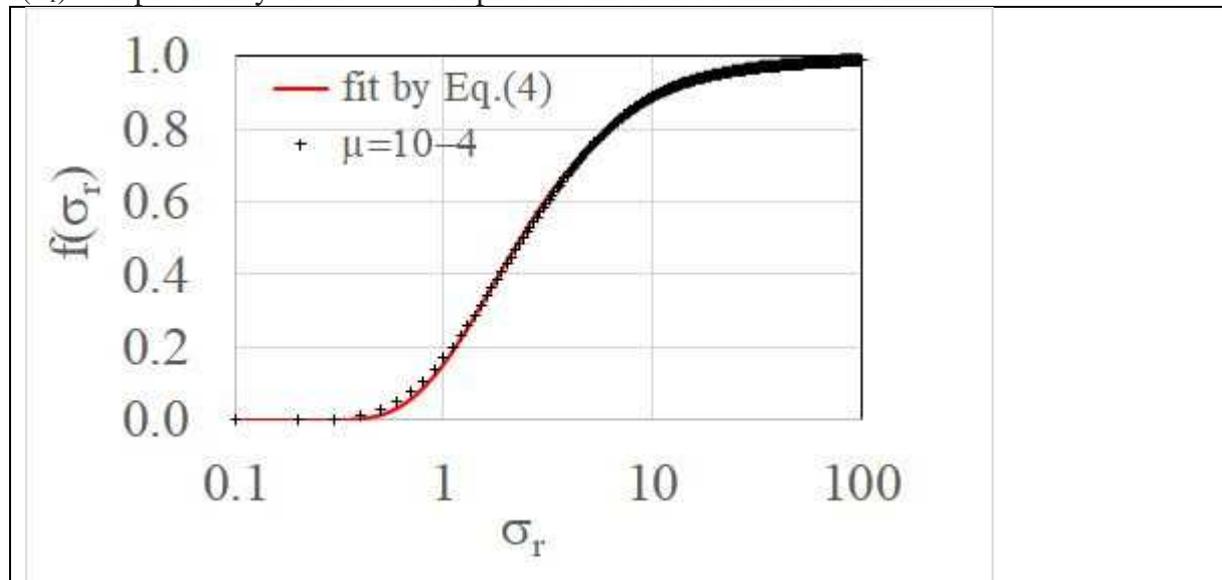

Fig.9 Fit of the values of $f(\sigma_r)$ for $\mu=10^{-4}$ (R. Radhakrishnan et al.[52]) with Eq.(8) and $\lambda=1.712$ ; q=1.163

It is also possible to use a prefactor in Eq.(8) $f_{max}(\Phi)$ [54], but at the cost of a supplementary parameter which will not be useful to interpret our experimental data.
In order to fit the dependence of the viscosity versus the volume fraction for different values of the friction coefficient Singh et al [28], have proposed to generalize Eq.(2) to take into account the variation of the viscosity with the friction coefficient. Their modification is the following:

$$\eta_c(\Phi, \sigma_r) = \alpha(\sigma_r, \mu)(1 - \frac{\Phi_{eff}}{\Phi_J(\sigma_r,\mu)})^{-2} \qquad (9)$$

Where the function α has the dimension of a viscosity with:

$$\alpha(\sigma_r, \mu) = \alpha_\mu f(\sigma_r) + \alpha_0(1 - f(\sigma_r)) \qquad (10)$$

and $\quad \Phi_J(\sigma_r, \mu) = \Phi_j^\mu f(\sigma_r) + \Phi_0(1 - f(\sigma_r)) \qquad (11)$

The parameters $\alpha_\mu$ and $\Phi_j^\mu$ are functions of the friction coefficient, μ, and are obtained from a fit of the numerical results with an empirical function. Wagner *et al*. [33] have used their prediction for these functions together with experimentally determined values of the friction coefficient in order to compare the predictions of the model with some experimental results on several kind of suspensions. They found strong deviations between the model and their experimental results.

In our system we know that the coating molecule is a superplasticizer playing the role of a polymer brush at the surface of the particles and we have seen that the divergence of the low stress viscosity (Fig.(5)) was corresponding to the random close packing of our suspension; it means that the friction coefficient is close to zero. This agrees with measurement made with a surface force balance on mica surfaces covered by PEO polymer of similar length where the friction coefficient was of order $10^{-3}$ [55,56]. In the model we need the lower jamming volume fraction $\Phi_j^\mu$. In principle it can be determined from the viscosity of the second branch of $\eta_c(\sigma_r)$ in the domain of DST where $\Phi < \Phi_j^\mu$ for large values of $\sigma_r$ where $f(\sigma_r) \to 1$ since in this case $\eta_c$ becomes a constant independent of the stress (cf Eqs (9)-(11)). In this zone -typically for φ ≤0.58 - there is a second branch, but due to high stresses and shear rates it is not possible to obtain a reproducible value of the viscosity (cf fig.4b). In the W-C model and in numerical simulations the friction coefficient μ is supposed to be independent of the shear stress but in practice and especially in the presence of a layer of adsorbed polymer at the surface of the particles, we expect that the friction increases with the stress due an increase of the entanglement between the polymers as already observed by AFM measurements where the friction increases a lot due either to the entanglement of the polymer or even to their expulsion from the surface [56,57]. In this context it is reasonable to suppose that $\Phi_j^\mu$ corresponds to the loose random packing at high friction (μ>1). For monodisperse spheres [52,58,59]: $\Phi_j^\mu \sim 0.55$-$0.56$ which is also close to the high friction value obtained by numerical simulation with the critical load model [28]. We can use these values instead of $\Phi_{RCP}$ in Eq.(2) for the transposition to our polydisperse suspension, and we obtain $\Phi_j^\mu \sim 0.592 - 0.602$. Using the second model based on the analogy with the bidisperse model we find $\Phi_j^\mu \sim 0.581 - 0.592$. Finally, to check the model we shall use the following values: $\Phi_0 = 0.684$ and $\Phi_j^\mu = 0.59$. In any event, we shall see that even a change of ±0.01 in $\Phi_j^\mu$ does not change our conclusions.

Besides these two volume fractions, the other parameters of the model are those defining the fraction of frictional contacts (Eq.(8). In a first step we could keep those used to fit the results of numerical simulations : λ=1.712; q=1.163 (cf.Fig.9 ), but the parameter λ is relative to a balance between the shear force and the repulsive force between 2 particles which is certainly not the same in our experiment as in simulation, so we shall take it as a fitting parameter. On the other hand it remains the two parameters $\alpha_0$ and $\alpha_\mu$ of Eq.11. The first one corresponds to the low stress case: $f(\sigma_r) \sim 0$ and corresponds to the beginning of the experimental curve $\sigma(\dot\gamma)$ which can be represented either by a Bingham model or more generally by a Herschel-Buckley model: $\sigma = \tau_y + K.\dot\gamma^p$. The second parameter $\alpha_\mu$ is a fitting parameter without real physical significance, so we shall discard it and take:

$$\alpha(\sigma_r, \mu) = \alpha_0 \qquad (12)$$

In our experiments, and quite generally in the presence of a yield stress, the beginning of the rheological curve is shear thinning. The yield stress can be either due to a primary aggregation caused by the attractive Van der Waals forces or on the contrary by repulsive forces producing a glassy state, but in systems where the range of the repulsive forces is important relatively to the radius of the particles, which is not our case. The combination of a yield stress and of discontinuous shear thickening has been analysed by numerical simulation in Singh *et al.* [60]. For moderate attractive force they find a shear thinning part at low stress followed by shear thickening and DST at higher stresses; for the highest attractive forces the DST was no longer observable. They found that the total viscosity was well represented by adding the Herchel-Buckley(HB) behavior at low stress and the contact viscosity (Eq.(9)) at high stress. In this approach, the beginning of the contribution of the contact viscosity depends strongly on the exponent p of the HB law. Since in the experiment we do not have access to the contact viscosity we adopt the view that we are not looking for a good model of the low stress behavior but rather to the application of the W-C model at intermediate and high stress. Then instead of the HB law we use the Bingham one (p=1) even if it does not fit very well the lower part of the stress versus shear rate curve. In this case we can write:

$$\sigma = \tau_y + (\eta_B + \eta_C(\sigma))\dot{\gamma} \quad \text{with} \quad \eta_c(\sigma) = \alpha_0 (1 - \frac{\Phi_{eff}}{\Phi_J(\sigma_r)})^{-2} \quad (13)$$

In this description $\eta_B$ is the plastic viscosity obtained by fitting the beginning of the curve by a Bingham law. and we suppose that all the shear thickening part is described by the W-C model represented by the contact viscosity $\eta_c$. On the other hand we have $\eta_c(\sigma \to 0) = \alpha_0(1 - \frac{\Phi}{\Phi_0})^{-2}$, then instead of subtracting this quantity to $\eta_c(\sigma)$ we can just incorporate $\eta_B$ in $\eta_c$ by imposing that the value of $\alpha_0$ gives back $\eta_B$ when $\Phi_J$ tends to $\Phi_0$ at low stress. Finally, we end up with:

$$\sigma = \tau_y + \eta_{pl}(\sigma)\dot{\gamma} \quad \text{with} \quad \eta_{pl}(\sigma) = \eta_B (1 - \frac{\Phi}{\Phi_0})^{+2}(1 - \frac{\Phi}{\Phi_J(f(\sigma_r))})^{-2} \quad (14)$$

In eq.(14) the only remaining parameters are those (λ,q) defining the function f($\sigma_r$) (cf Eq.(8)). They can be determined directly by the condition that the theoretical curve should pass through the transition point ($\sigma_c$, $\dot{\gamma}_c$).:

$$\sigma_c - \tau_y = \eta_{pl}(\sigma_c)\dot{\gamma}_c \quad \text{and} \quad \left.\frac{d\dot{\gamma}}{d\sigma}\right|_{\sigma=\sigma_c} = 0 \quad (15)$$

Or equivalently, using $\dot{\gamma}_c = (\sigma_c - \tau_y)/\eta_{pl}(\sigma_c)$, where $\eta_{pl}(\sigma)$ is given by Eq.(14) we end up with the two equations:

$$\frac{\sigma_c - \tau_y}{\dot{\gamma}_c} = \eta_{pl}(\sigma_r = 1) \quad (16)$$

$$1 = \frac{\sigma_c - \tau_y}{\eta(\sigma_c)}\left[\frac{d\eta_{pl}(\sigma)}{d\sigma}\bigg|_{\sigma=\sigma_c}\right] \quad (17)$$

The results are shown in Figs 10-12 for the volume fractions Φ=0.58, Φ=0.64 and Φ=0.66. The parameters used to obtain these curves are listed in table 1. The two first parameters are those of the Bingham law representing the part of the curve which is not shear thickening. The third one $\alpha_0$ is obtained from $\eta_B$ (Eq.(14)), then λ and q characterize the function f(σ) and the last quantity f($\sigma_c$) is the proportion of frictional contacts at the critical stress as obtained from the values of λ and q reported in Eq.(8) for $\sigma_r$=1. For Φ=0.58 the experiment was done in cylindrical Couette rheometry whereas for Φ=0.64 and Φ=0.66 the experiments were done in plate-plate geometries. For this last geometry the experimental curves presented in Figs (10)-(11) have been corrected as described by Eq.(4) and illustrated in Fig.(2).In fig.11 we have presented the results for Φ=0.58. The red curve is the experimental one, the black one is the theoretical one (obtained from a fit of the beginning of the curve ($\dot{\gamma} < 200 s^{-1}$) by a Bingham law, The resulting curve does not fit the experimental one very well below the critical point and not at all above. The turning point corresponding to the S shape (not represented here) is found at $\sigma \sim 6000 Pa$ instead of about 1600Pa experimentally. It is worth noting that adding the parameter $\alpha_\mu$ (Eq.(10) in the prefactor of the viscosity) does not improve significantly the agreement between the experiment and the model. On the other hand, this discrepancy can't be attributed to the uncertainty on $\Phi_j^\mu$ as can be seen on Fig.10 where the two curves with $\Phi_j^\mu = 0.58$ and $\Phi_j^\mu = 0.60$ are still far from the experimental one.

| Φ | τy(Pa) | ηB | α0 | λ | q | f(σc) |
|---|---|---|---|---|---|---|
| 0.58 | 2.1 | 0.81 | 0.018 | 1.08 | 1.19 | 0.335 |
| 0.64 | 18.4 | 4.92 | 0.020 | 1.46 | 3.50 | 0.023 |
| 0.66 | 40.1 | 28.5 | 0.034 | 1.02 | 111 | 3.3 10$^{-4}$ |

Table 1 Parameters used in the Wyart Cates model for the volume fractions Φ=0.58, 0.64, 0.66

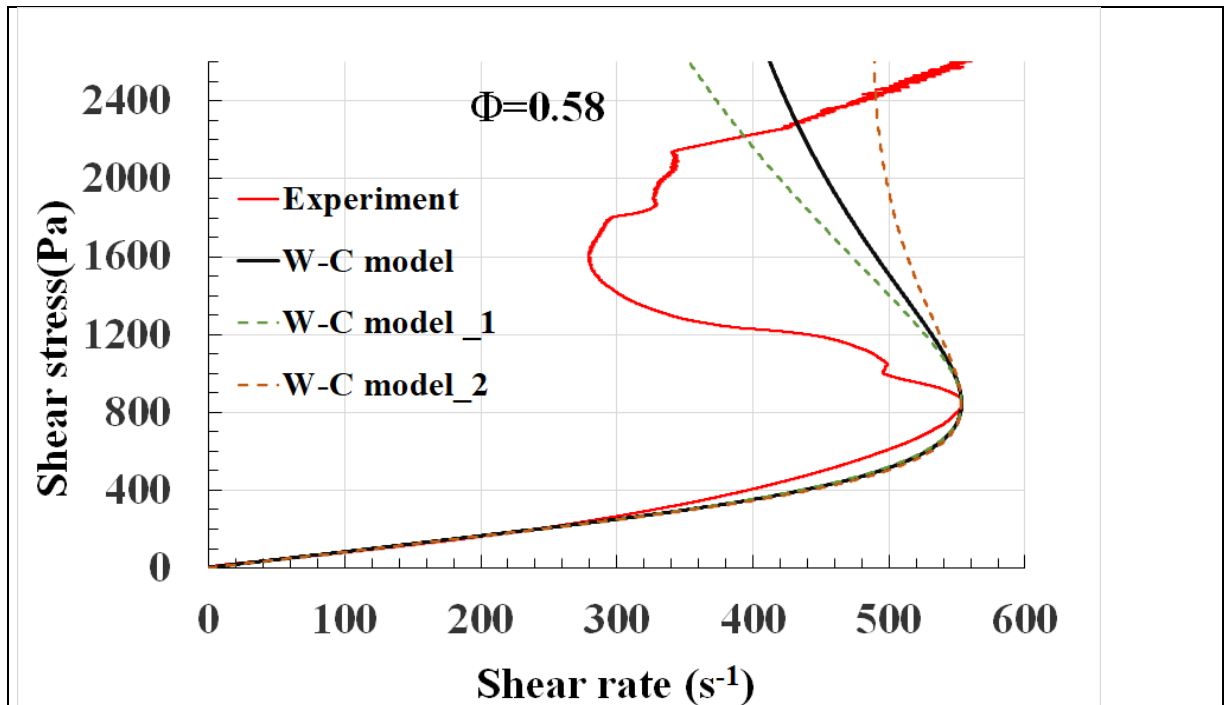

Fig10 Φ=0.58 Comparison of the experimental curve (in red) to the theoretical one(in black) with $\Phi_j^\mu$. =0.59. The dashed curve in green corresponds to $\Phi_j^\mu$. =0.58 and the one in brown to $\Phi_j^\mu$. =0.60

The results for the volume fractions Φ=0.64 and Φ=0.66 are presented in Figs.11a and 11 b. For the moment we discard the dotted lines which will be discussed later. For Φ=0.64 we see in the linear scale that we are close to a pure Bingham behavior but, there is a small shear thickening before the transition. The value q=3.50 (cf Table 1) indicates that the transition is more abrupt than at Φ=0.58 where q=1.19-(close to the value q=1.16 of the numerical simulation).

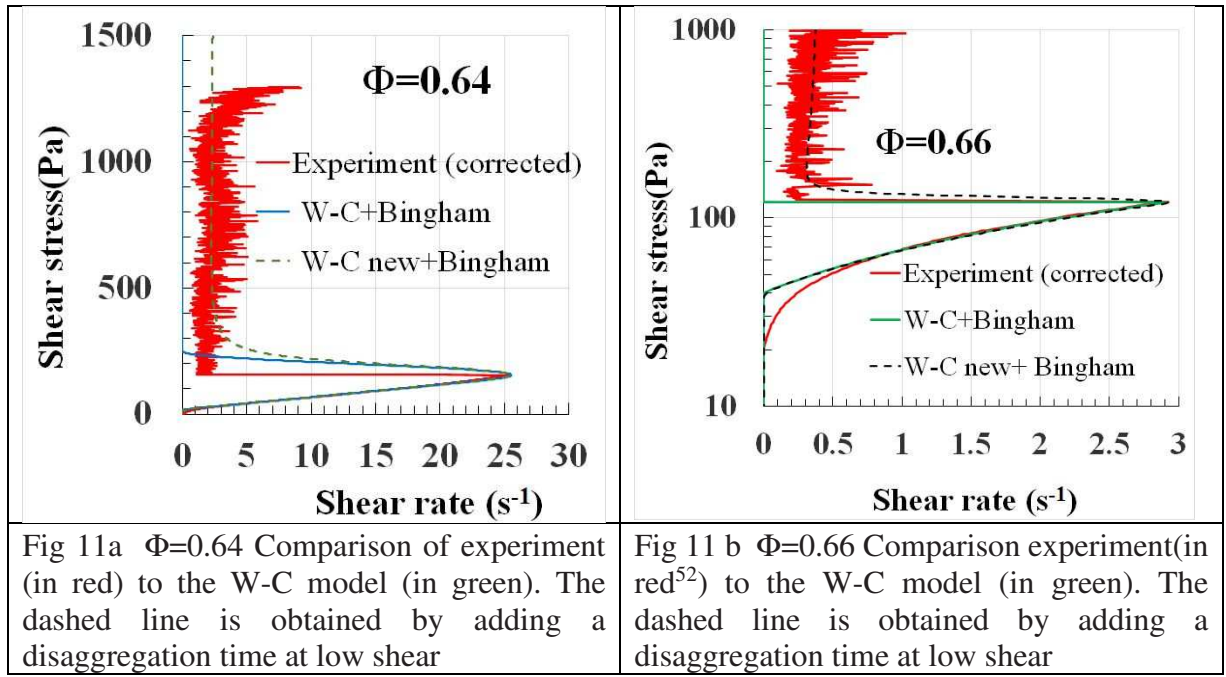

| Fig 11a Φ=0.64 Comparison of experiment (in red) to the W-C model (in green). The dashed line is obtained by adding a disaggregation time at low shear | Fig 11 b Φ=0.66 Comparison experiment(in red[52]) to the W-C model (in green). The dashed line is obtained by adding a disaggregation time at low shear |
|---|---|

On the contrary at Φ=0.66 we have a transition which occurs without being preceded by shear thickening. A fit of the upper part of the curve by a Bingham law well represents the experimental behavior above 1 s$^{-1}$ (cf Fig.12b). As there is no shear thickening before the transition, it amounts to say that the percolating network of frictional contacts is created suddenly at $\sigma=\sigma_c$, what is reflected by the huge value q=111. In other words the function $f(\sigma_r)$ jumps from 0 to 1 at $\sigma=\sigma_c$ We have plotted in Fig.12 the evolution of this function with the stress. We see that at Φ=0.58, the progressive increase of $f(\sigma_r)$ is similar to the one observed in numerical simulation but with a shift relatively to $\sigma_r$. This shift is not surprising since the value of λ depends on the specific shape of the repulsive barrier preventing the particles to come into contact. The fact that there is only a small shear thickening before the transition at Φ=0.64 is translated by a sharper evolution of the function $f(\sigma_r)$, and finally we have the step function at Φ=0.66.

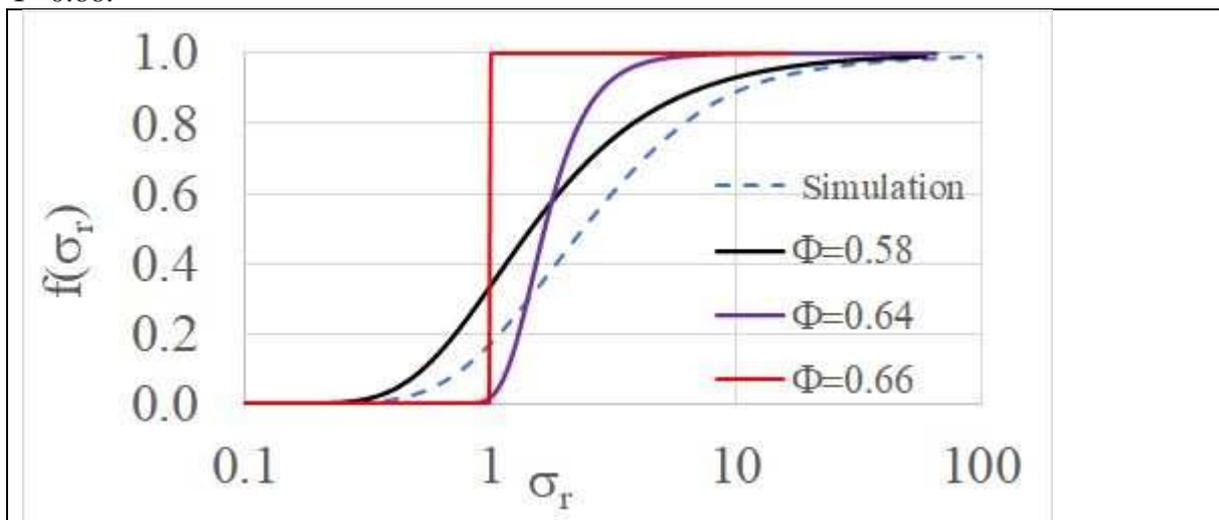

Fig.12 Evolution of the fraction of frictional contacts for different volume fractions with the relative stress $\sigma_r=\sigma/\sigma_c$ The dashed line refers to the result of numerical simulation which does not depend on volume fraction (Radhakrishnan et al.[52], Mari et al.[27])

This evolution of f(σ) with the volume fraction is supported by the measurement of the resistance versus the stress (Figs 7-8) which shows qualitatively the same difference of behavior between the volume fraction Φ=0.64 where the drop of resistance is abrupt and Φ=0.55 where it is much more progressive. It is also worth pointing that this drop of resistance at the DST transition is an experimental signature of the formation of a percolating network of frictional contacts. The fact that, contrary to the simulation results, the fraction of frictional contacts for the same stress depends strongly on the volume fraction is likely due to the complexity of the interaction between the layers of adsorbed polymers which react differently to an homogeneous increase of pressure due to an increase of volume fraction and to an anisotropic increase of pressure due to the shear stress. In other words, for the same shear stress, the interpenetrating zone is higher at high volume fraction and the applied shear stress will be more efficient to remove the layer of adsorbed polymer. In a recent paper Guy *et al.* [61] using the simulation of a bidisperse suspension with a ratio of four in the diameters, have shown that the slope of f(σ) can be lower for certain (20%) proportion of small spheres. In our case the polydispersity is moderate (<a>/$\sigma_{std}$ =0.5) and above all it is the change of f(σ) with the volume fraction that we follow and not a change of composition of the suspension.

## C. The absence of jamming above the DST transition

In the W-C model the suspension should stop to flow above a given stress if the volume fraction is between $\Phi_0$ and $\Phi_j^\mu$ because in this range the jamming volume fraction $\Phi_j(\sigma)$ will always reach the actual volume fraction Φ when the stress increases, causing the divergence of the viscosity. In practice we see (cf figs 11a, 11b) that, instead of going to zero the shear rate keeps, in average, an almost constant value when the stress is increased above the DST transition. As pointed out when discussing the results presented in Fig.(4) many authors already noticed this kind of behavior on different kinds of suspensions. One could object that this residual shear rate is due to some slipping of the suspension on the walls, but on the other hand if we apply a magnetic field of about 100kA/m to the same suspension of carbonyl iron particles, it will show a yield stress of several kPa without any slipping in the same plate-plate geometry [62,63]. Besides, the role of the inertia of the rotating tool was recognized to play a major role in the instability which occurs when the differential viscosity is negative [42,44,64,65]. When coupled to the W-C model the introduction of an exponential relaxation for the time evolution of f($\sigma_r$) allows to well recover the oscillations above the transition but did not explain the persistence of these oscillations at stresses where the W-C model predicts the total stop of the flow. By adding in the time evolution of f($\sigma_r$) a second term: H($f_m$-f) allowing the growth of f with the shear rate it is possible to obtain an asymptotic value of the shear rate at high shear stress [66] but at the condition to take $H \propto \sigma^{3/2}$ which seems rather arbitrary. In a recent paper on DST in capillary flow [67] we have proposed another explanation and a modification of the W-C model which well succeeded to reproduce this non-zero -almost constant- shear rate at high stresses. Our approach is based on the idea that a state of flow arrest at high stress which would be only due to friction and not to adhesive forces, should be unstable. This is because, in the absence of flow and of large enough adhesive forces, the entropic forces - like those due to a small residual Brownian motion or to a change of configuration of the coating molecules present on or around the surfaces-will be enough to destroy some fragile links in the network of frictional contacts and will lead to a restart of the flow. This mechanism will give, on average, a non-zero shear flow at high stresses, and can be simply taken into account by inserting the condition that the fraction of frictional contacts should vanish if the shear rate tends to zero, whatever the value of the stress. This can be done for instance by multiplying the function f($\sigma_r$) (Eq.(8)) by a function ,g, of the shear rate which vanishes at zero shear rate and tends to unity when it increases:

$$f'(\sigma_r, \dot{\gamma}) = f(\sigma_r) * g(t_d.\dot{\gamma}) \tag{17}$$

where $t_d$ is a parameter related to the relaxation time of the shear rate. We previously took arbitrarily a Langevin function for g[67], but the precise shape of this function can be deduced from an evolution equation of the function $f(\sigma_r)$ which is a structural parameter like others used to describe the time dependent rheology. Such an approach combines two mechanisms, one for the building of the structure and the other for its destruction [2]. In this way, we will write:

$$\frac{\partial f}{\partial t} = -\frac{1}{t_B}\left(f(t) - f(\sigma_r)\right) - \frac{f(t)}{t_d} \tag{18}$$

The first term describes the relaxation of the structure to its equilibrium value and was already used to reproduce the oscillations of the shear rate above the transition [42]. If the fraction of frictional contacts is below its equilibrium value, it will rise with a characteristic time $t_B$ which will be inversely proportional to the collision rate and so to the shear rate: $t_B = 1/(\dot{\gamma} C_B)$ where $C_B$ is a constant. On the other hand, in the absence of shear, we expect that the frictional contacts will be destroyed with a characteristic time $t_d$ that, for simplicity, we suppose independent of the stress. At equilibrium $\frac{\partial f}{\partial t} = 0$ and from Eq.(18) we have:

$$f_{eq}(\sigma_r) \equiv f'(\sigma_r, \dot{\gamma}) = f(\sigma_r)\frac{\dot{\gamma}}{\frac{1}{t_d.C_B}+\dot{\gamma}} = f(\sigma_r)g(\dot{\gamma}) \tag{19}$$

Now the jamming volume fraction depends also on the shear rate since f is replaced by f' in Eq.(11). At high value of $\sigma_r$ we have $f(\sigma_r) \to 1$ then $\Phi_j \to (1-g(\dot{\gamma}))\Phi_0 + g(\dot{\gamma})\Phi_m$ and we end up with:

$$\eta \to \frac{1}{(\Phi_j-\Phi)^2} = \frac{1}{(\dot{\gamma}-\dot{\gamma}_\infty)^2} \quad \text{with} \quad \dot{\gamma}_\infty = \frac{1}{C_B.t_d}\left(\frac{\Phi_0-\Phi}{\Phi-\Phi_m}\right) \tag{20}$$

The asymptotic value of the shear rate $\dot{\gamma}_\infty$ is given by Eq.(20). The dotted curve of Fig. (11a) for $\Phi=0.64$ is obtained with $C_B.t_d=0.3$s, giving $\dot{\gamma}_\infty = 2.73\ s^{-1}$ and the one of Fig(11b) for $\Phi=0.66$ with $C_B.t_d=0.7$s. giving $\dot{\gamma}_\infty = 0.43\ s^{-1}$. Note that it is the product $C_B.t_d$ which can be obtained from the fit of the experimental curve and not each parameter separately. With this modification of the W-C model we have now a good agreement with the experiment. Still we are using the equilibrium value $f_{eq}(\sigma_r)$ and not its time evolution described by Eq.(18). It is only by considering this time evolution together with the introduction of the inertia of the rotating part that we can describe the fluctuations of the shear rate above the DST transition and obtain separately the parameters $t_d$ and $C_B$. This is beyond the scope of this paper but we have verified that with the use of Eq.(18) the oscillating regime continue above the jamming volume fraction predicted by the W-C model.

In practice, above a few kPa, due to the centrifugal force and also to the presence of a normal force $\sigma_{rr}$, transmitted through the percolated network, the interface with air becomes irregular with extrusion of "granules" and intrusion of air bubbles. This mechanism exists as well in plate-plate geometry and in cylindrical Couette geometry [25] and prevents to get reliable results at very high stresses even in cylindrical Couette geometry.

Before passing to the effect of the magnetic field, from these comparisons between experiments at three typical volume fractions we can already retain the following conclusions:
1) from the measurement of the electric resistance of the suspension it is possible to follow the formation of the network of frictional contacts when increasing the stress.
2)-It is not possible to represent the rheological curve at every volume fraction with the same values of the function $f(\sigma_r)$: if the parameter q=1.16 (obtained in some numerical simulations) of the stretched exponential can describe the curve at $\Phi=0.58$, this is clearly not the case at $\Phi=0.64$ and at $\Phi=0.66$ there is no shear thickening at all before the transition, leading to a Heaviside function. It is the expression of the fact that we pass abruptly from a Bingham or even a shear thinning behavior (p<1) to the DST transition.

2) The non-zero average shear rate at high stress is explained by taking into account that the frictional contacts are unstable at zero shear rate.

## IV. EFFECT OF THE MAGNETIC FIELD ON THE DST TRANSITION

We have shown in preceding papers [36,68] that the application of a magnetic field on a suspension of carbonyl iron particles at high volume fraction could considerably shift the critical shear rate of the DST transition towards lower values. We also remarked that the difference between the critical stress and the yield stress remained approximately constant in the range of field we have used [64]. We shall try in this section to understand the physical process which could explain this behavior. In the following figures (Fig.13a, 13b) we have plotted the raw curves obtained in plate plate geometry and below the differential viscosity $\frac{\partial \sigma}{\partial \dot{\gamma}}$ determined from the corrected curve as defined by Eq.(4)

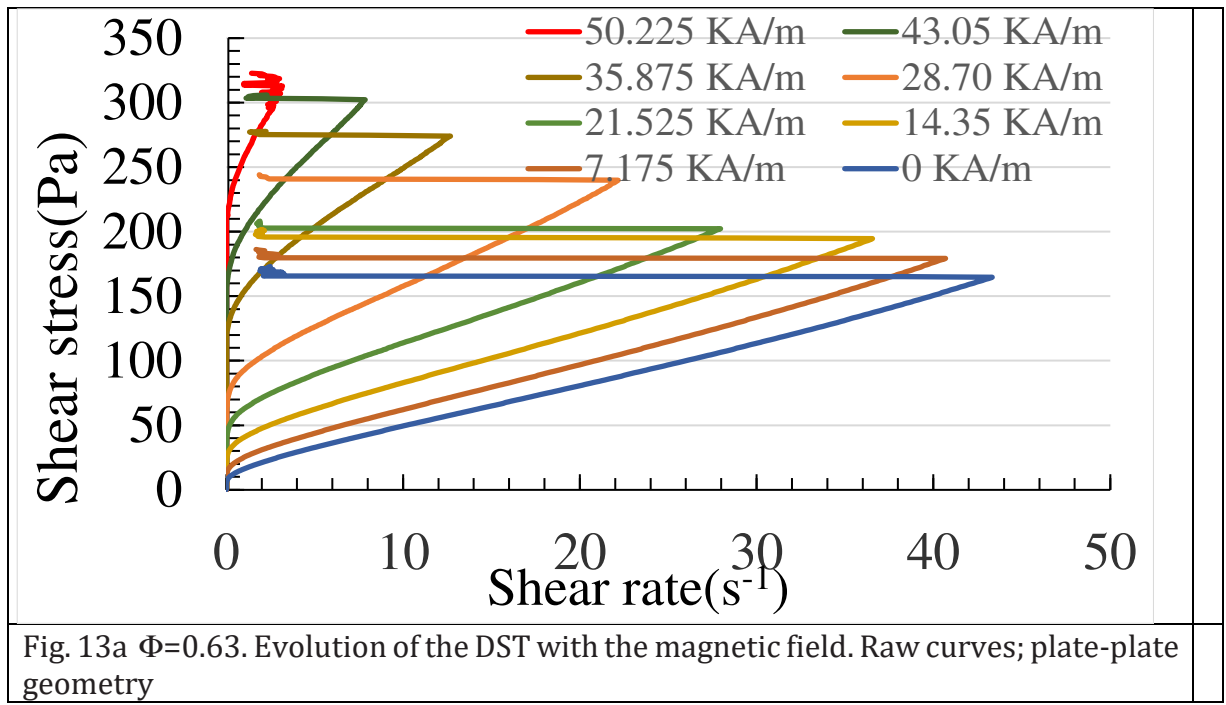

Fig. 13a Φ=0.63. Evolution of the DST with the magnetic field. Raw curves; plate-plate geometry

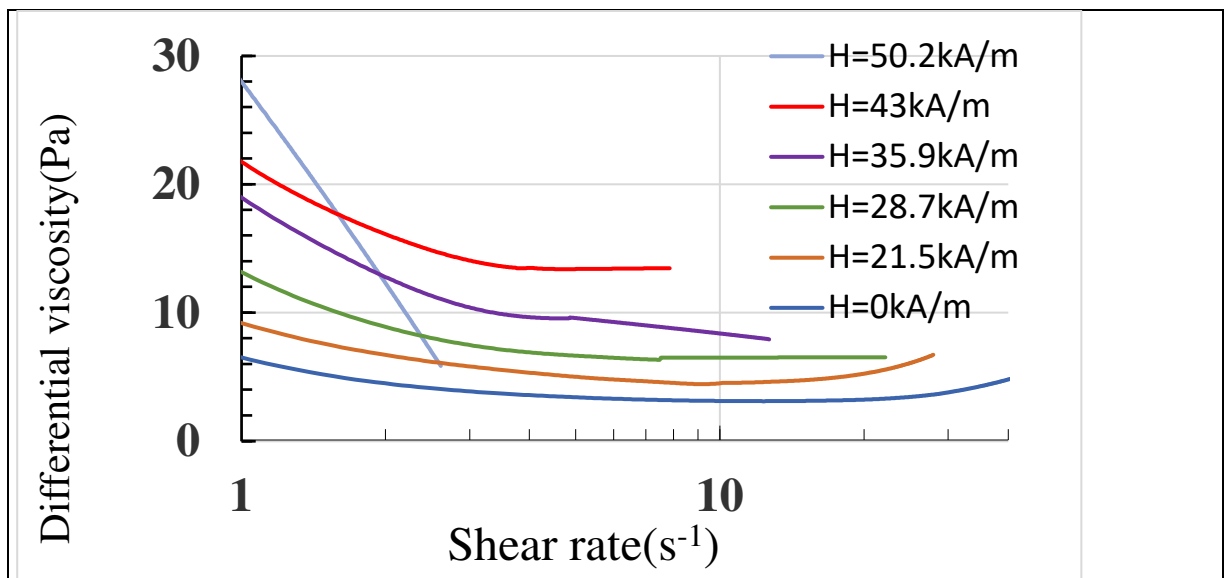

Fig. 13b Φ=0.63. Differential viscosity for some of the rheograms represented in Fig.13a

The more remarkable result is that, increasing the magnetic field, we pass from a behavior which is first shear thinning and then shear thickening to a behavior where we have only shear thinning before the DST transition. This is similar to what we observe in the absence of the field when we increase the volume fraction. Once again, in the frame of the W-C model, this is only possible if the function $f(s_r)$ rises suddenly from zero to one at the transition (cf Fig. 12). The shear thickening part which precedes the DST transition at low field or low volume fraction is more usual and can be interpreted as the formation of small clusters due to frictional contacts provoked at high shear by the removal of the layer of polymer from the surface [56]. Some more information on the role of the polymer layer can be inferred from the comparison of the evolution of the yield stress and of the viscosity with the magnetic field at intermediate and high volume fraction. This is shown respectively in Figs 14a and 14b.

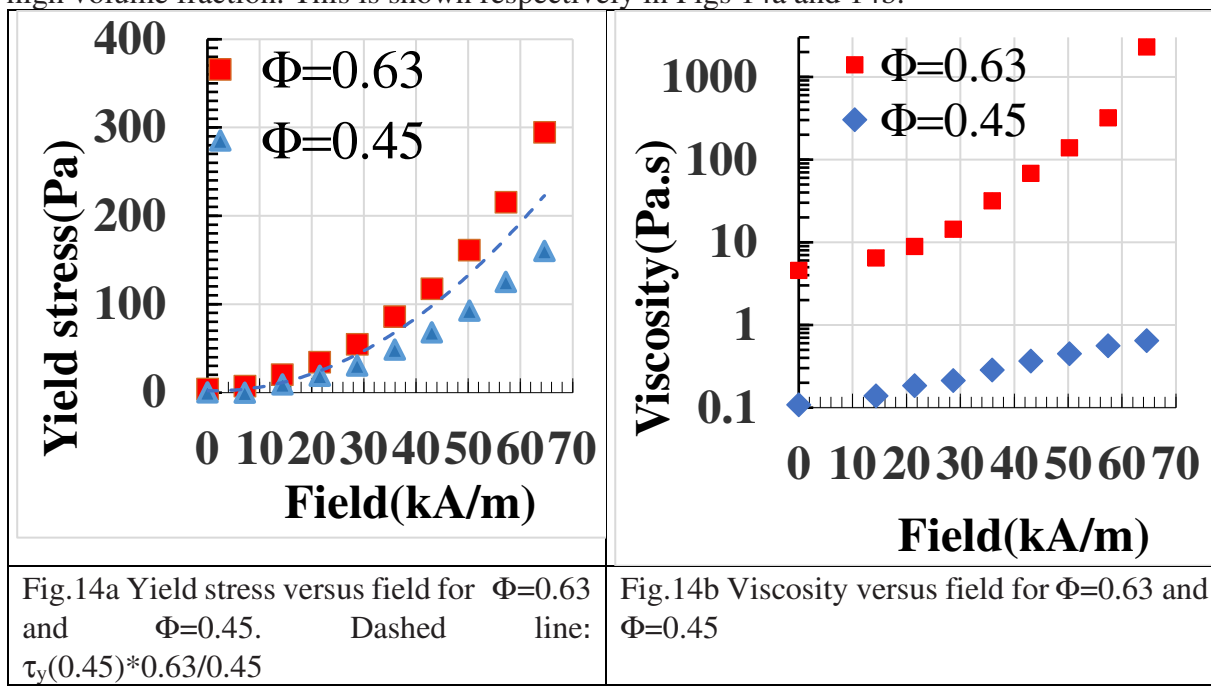

Fig.14a Yield stress versus field for Φ=0.63 and Φ=0.45. Dashed line: $\tau_y(0.45)*0.63/0.45$

Fig.14b Viscosity versus field for Φ=0.63 and Φ=0.45

For the yield stress, the difference between the two volume fractions is not important. Actually, if we consider the simplified model of independent chains of particles spanning the gap between the two plates, the yield stress should be just proportional to the number of chains, that is to say, to the volume fraction; the dashed line is the extrapolation for Φ=0.63 from Φ=0.45 if it was the case. Even if the experimental values grow slightly faster (red squares), the difference is not so big and easy to explain since the model of individual chains is no longer valid at high volume fractions since the number of contacts per particles- and so the magnetic attractive force between pairs of particles- is expected to increase with the volume fraction. On the contrary the increase of viscosity with the intensity of the magnetic field is about 3 orders of magnitude at F=0.63 against one order of magnitude at Φ=0.45. The viscosity in Fig.14b is an average plastic viscosity: $\eta_{63} = (\sigma_c - \tau_y)/\dot{\gamma}_c$ for Φ=0.63 and $\eta_{45} = (\sigma(400s^{-1}) - \tau_y)/400$ for Φ=0.45. We must emphasize that the range of applied stress is about the same at Φ=0.63 as at Φ=0.45 but it is the range of shear rate which is much larger at Φ=0.45 than at Φ=0.63.

This unexpected large increase of the viscosity with the field is an important observation for two reasons; firstly it will allow to trigger the viscosity of this magnetorheological fluid with a much larger efficiency than with usual ones based on suspensions of intermediate volume fractions and secondly it helps to understand the process leading to DST in the presence of polymer brushes. The interaction between two layers of polymer brushes has been extensively studied [69,70] mainly because of their applications to reduce the viscosity of concentrated suspensions of mineral particles as for instance in cement industry. The repulsive force of entropic origin between the tails of the polymers prevents the particles from aggregation in the presence of attractive Van Der Waals forces. This osmotic force depends on many factors like the size distribution and the conformation of the polymer its energy and density of adsorption; its miscibility with the suspending fluid (the Flory parameter). A key parameter which derives from these characteristics is the interpenetration zone of width ,d, of the layers of polymer and its dependence on the applied hydrodynamic (and magnetic in our case)stress . If d remains small, then the system is equivalent to a hard sphere suspension with a renormalized volume fraction, F$_{eff}$-incorporating the thickness of the polymer layer and a short-range repulsive force. If the interparticle force, generated by the externally imposed stress, increases, the value of d will also increase in an extent depending of the stiffness of the repulsive force and also of a Weissenberg number $W = \tau.\dot{\gamma}$ where t is the relaxation time of an adsorbed polymer. If *W<1* the polymer has the time to recover its equilibrium shape between the next collision with another polymer; then the shear force generated by the collisions between monomers in the interpenetration zone grows proportionally to the shear rate[71]. It means that in this regime we can have a Newtonian behavior but with a viscosity which will depend on the field since increasing the field will increase the interpenetration zone. On the contrary if *W>1* the polymers do not have time to relax: they remain stretched by the shear flow which will decrease the interpenetration zone and give a shear thinning behavior or even an absence of dependence of the shear if the interpenetration zone remains very weak. The relaxation time of the polymer can be estimated from the Rouse model for PEO polymer in water [72]: $\tau = 0.0142N^2\xi b^2$/kT. With N=44 the number of units of PEO, x the friction coefficient on one unit and b the length of one unit. Considering that the viscosity of our suspending fluid is 10 times the one of water we get $\tau \sim 10^{-6}s$. Another estimation based on Zimm theory [73] gives $\tau = 5.11\ R^3\eta_s$/kT with *R* the root mean square separation of the extremities of a polymer and $\eta_s = 0.011 Pas$, the viscosity of the suspending fluid. Taking into account the expansion of the polymer due to its compression by its neighbours we have R=6nm [42] and in this case τ=2.6 10$^{-6}$s. This order of magnitude means that, with this small polymer, we shall always have W<<1 and so that we should remain in a Bingham regime with a plastic viscosity independent of the shear rate. This is roughly what we observe above 3 s$^{-1}$ except for the highest field where we have a continuous shear thinning (cf. Fig 13b). This change of regime at high compression could be due to a structural change of the compressed layer perhaps related to a beginning of desorption of the polymer before the DST transition. On the other hand, the shear thinning observed at $\dot{\gamma} < 3s^{-1}$ is simply due to the progressive rupture of the aggregates formed by the attractive magnetic forces.

If we call $\eta_b$ the viscosity of these bilayers of polymers which separate the surfaces of the particles it is likely that, at high volume fraction where the polymer layers are always interpenetrated, we could approximate the total viscosity as $\eta=\eta_{HS}.\eta_b$ where $\eta_{HS}$ is the relative viscosity of the hard sphere suspension without coating polymers. Finally, increasing the field increases a lot the plastic viscosity and to a much less extent the yield stress - which is the stress necessary to separate the particles against the attractive magnetic force. On the other hand, the stress needed to sweep the polymer layer out of the surface in the lubricated zone is essentially the shearing stress coming from the relative motion of the particles. This is the reason why it is the difference between the critical stress and the yield stress which remains approximately constant when we increase the magnetic field instead of the critical stress as we can see in the following figures.

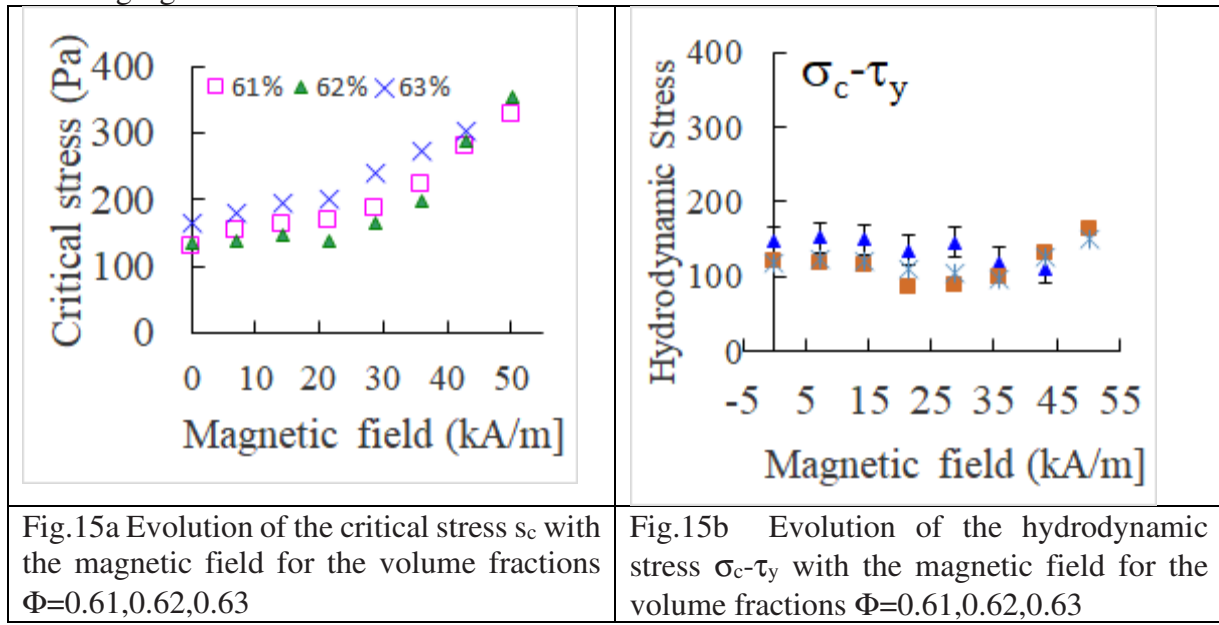

| Fig.15a Evolution of the critical stress $s_c$ with the magnetic field for the volume fractions $\Phi=0.61, 0.62, 0.63$ | Fig.15b Evolution of the hydrodynamic stress $\sigma_c-\tau_y$ with the magnetic field for the volume fractions $\Phi=0.61, 0.62, 0.63$ |
|---|---|

The attractive force induced by the magnetic field contributes to increase the interpenetration of the polymer and consequently the viscosity related to the shearing forces between the interpenetrated parts of the polymer increases, until the shearing stress needed to wipe the polymer out of the surface is reached. In this process the yield stress generated by the magnetic force between the iron particles does not contribute directly to the shearing force acting on the polymer layer which is responsible for the desorption of the polymer and the DST transition.

## V. CONCLUSION

Using a suspension of ferromagnetic particles stabilized by a superplasticizer molecule used in cement industry, we have obtained a discontinuous shear thickening in a broad range of volume fraction ($0.54<\Phi<0.67$). From the size distribution of the particles, we have deduced the two volume fractions $\Phi_0$ and $\Phi_j^\mu$ on which are based the Wyart-Cates model of DST. This model also introduces the function $f(\sigma_r)$ representing the fraction of frictional contact versus the stress; we have modelled this function with the help of two parameters $\lambda$ and $q$; this last one depicting the sharpness of the transition. These two parameters are obtained from the constraint that the experimental curve passes through the critical point where the shear rate begins to decrease. Whereas simulations predict that this function remains independent of the volume fraction, we find that it fits the one obtained in simulation, only in the domain of soft DST transition at the lowest volume fractions where there is a second regime of constant viscosity. At highest volume fraction the transition is sharper and finally becomes steplike. This last behavior is related to the fact that there is no shear thickening before the transition, which implies that $f(\sigma_r)=0$ for $\sigma_r<1$. The measurement of the electric resistance together with the stress/shear rate curve allows to confirm the onset of a percolated network of frictional contacts associated to the decrease of the shear rate and also the fact that the expansion of this network with the stress depends strongly on the volume fraction, contrary to the predictions of the numerical simulations. A numerical model introducing a repulsive force depending on the interpenetration of the polymer together with a criteria for the desorption of the polymer should allow to recover this behavior.

Contrary to the W-C model which predicts the existence of a domain of jamming above the DST transition, we do not observe it experimentally but rather the shear rate remains, on average, constant above the critical stress. We were able to reproduce this behavior by introducing a relaxation time, $t_d$, of the frictional contacts at zero shear rate expressed in Eq.(18). The resulting asymptotic shear rate $\dot{\gamma}_\infty$(Eq.(20)) gives access to the product $C_B.t_d$ where $C_B$ is a constant associated to the collision rate between particles. The independent determination of these two parameters could be done through the analysis of the period of oscillations of the shear rate above the transition.

We have explained qualitatively the evolution of the rheology in the presence of the magnetic field by its effect on the interpenetration of the polymer layer adsorbed on the surface of the particles, in particular the strong increase of plastic viscosity for small amplitudes of magnetic field. The evolution of the polymer layers with their progressive interpenetration and their desorption above a given shearing stress makes the particle short range interactions much more complicated than the "critical load model" CLM model used in computer simulations, so it is not surprising that some predictions of these numerical models do not apply to our experimental systems. In order to get more information on the interparticle forces in the presence of brush polymer it would be useful to make experiments on a pair of iron microparticles with the help of a force apparatus [57]. At last, we want to emphasize that this magnetorheological fluid, based on very high volume fraction of iron particles thanks to this superplasticizer, is much more efficient than usual ones because of two physical phenomena: the DST transition and the increase of viscosity due to the interpenetration of the polymer brushes. The increase of yield stress with the field which is the usual mechanism in conventional MR fluid is of course present but not more than in usual MR fluids.

## ACKNOLEWGMENTS

The authors want to thank the CENTRE NATIONAL D'ETUDES SPATIALES (CNES, the French Space Agency) for having supported this research.